\documentclass[prl,showpacs,twocolumn]{revtex4}%
\input{psfig.sty}
\usepackage{amsmath}
\usepackage{graphicx}
\usepackage{amsfonts}
\usepackage{amssymb}%

\begin{document}
\title{Quantum telecommunication based on atomic cascade transitions}
\date{\today }
\author{T. Chaneli\`{e}re, D. N. Matsukevich, S. D. Jenkins, T. A. B. Kennedy, M. S. Chapman, and A. Kuzmich}
\affiliation{School of Physics, Georgia Institute of Technology, Atlanta, Georgia 30332-0430} 
\begin{abstract}
A quantum repeater at telecommunications wavelengths with long-lived atomic memory is proposed, and its critical elements are experimentally
demonstrated using a cold atomic ensemble. Via atomic cascade emission, an entangled pair of 1.53 $\mu$m and 780 nm photons is generated. The
former is ideal for long-distance quantum communication, and the latter is naturally suited for mapping to a long-lived atomic memory. Together
with our previous demonstration of photonic-to-atomic qubit conversion, both of the essential elements for the proposed telecommunications
quantum repeater have now been realized.
\end{abstract}
\maketitle

A quantum network would use the resources of distributed quantum mechanical entanglement, thus far largely untapped, for the communication and
processing of information via qubits \cite{briegel,duan}. Significant advances in the generation, distribution, and storage of qubit entanglement
have been made using laser manipulation of atomic ensembles, including atom-photon entanglement and matter-light qubit conversion
\cite{matsukevich}, Bell inequality violation between a collective atomic qubit and a photon \cite{matsukevich1}, and light-matter qubit
conversion and entanglement of remote atomic qubits \cite{matsukevich2}. In each of these works photonic qubits were generated in the
near-infrared spectral region. In related developments, entanglement of an ultraviolet photon with a trapped ion \cite{blinov} and of a
near-infrared photon with a single trapped atom \cite{weber} have been demonstrated. Heterogeneous quantum network schemes that combine
single-atom and collective atomic qubits are being actively pursued \cite{saffman}. However, photons in the ultraviolet to the near-infrared
range are not suited for long-distance transmission over optical fibers due to high losses.

In this Letter, we propose a telecommunications wavelength quantum repeater based on cascade atomic transitions in either (1) a single atom or
(2) an atomic ensemble. We will first discuss the latter case, with particular reference to alkali metals. Such ensembles, with long lived ground
level coherences can be prepared in either solid \cite{arndt} or gas \cite{matsukevich1} phase. For concreteness, we consider a cold atomic vapor
confined in high-vacuum. The cascade transitions may be chosen so that the photon (signal) emitted on the upper arm has telecommunication range
wavelength, while the second photon (idler), emitted to the atomic ground state, is naturally suited for mapping into atomic memory.
Experimentally, we demonstrate phase-matched cascade emission in an ensemble of cold rubidium atoms using two different cascades: (a) at the
signal wavelength $\lambda _s =776$ nm, via the $5s_{1/2} \rightarrow 5d_{5/2}$ two-photon excitation, (b) at $\lambda _s =1.53$ $\mu$m, via the
$5s_{1/2} \rightarrow 4d_{5/2}$ two-photon excitation. We observe polarization entanglement of the emitted photon pairs and superradiant temporal
profiles of the idler field in both cases.

We now describe our approach in detail and at the end we will summarize an alternative protocol for single atoms. {\it Step (A)} - As illustrated
in Fig. 1(a), the atomic sample is prepared in level $|a\rangle $, e.g., by means of optical pumping. It is important to note that, in the case
of an atomic ensemble qubit, an incoherent mixture of Zeeman states is sufficient for our realization. The upper level $|d\rangle $, which may be
of either $s$- or $d$-type, can be excited either by one- or two-photon transitions, the latter through an intermediate level $|c \rangle $. The
advantage of two photon excitation is that it allows for non-collinear phase matching of signal and idler photons; single photon excitation is
forbidden in electric dipole approximation and phase-matched emission is restricted to a collinear geometry (this argument implicitly assumes
that the refractive index of the vapor is approximately unity). Ideally the excitation is two-photon detuned from the upper level $|d \rangle $,
creating a virtual excitation.

{\it Step (B)} - Scattering via the upper level $|d \rangle $ to ground level $|a \rangle $ through the intermediate level $|e \rangle $  (where
$|e \rangle $ may coincide with $|c \rangle $) results in the cascaded emission of signal and idler fields. The signal field, which is emitted on
the upper arm, has a temporal profile identical to that of the laser excitation as a consequence of the large two photon detuning. As noted
earlier, the wavelength of this field lies in the 1.1-1.6 $\mu$m range, depending on the alkali metal transition used. The signal field can be
coupled to an optical fiber (which may have losses as low as 0.2 dB/km) and transmitted to a remote location.

The temporal profile of the idler field can be much shorter than the single-atom spontaneous decay time $t_s$  of the intermediate level. Under
the conditions of a large Fresnel number of the exciting laser fields, the decay time is of order $t_s /d_{th}$, characteristic of superradiance
\cite{eberly,mandel}. Here $d_{th} \approx 3n\lambda^2  l/(8\pi)$ is the optical thickness, where $\lambda $ is the wavelength, $n$ is the number
density and $l$ is the length of the sample.

\begin{figure}[htp]
\vspace{-0.3cm}
\begin{center}
\leavevmode  \psfig{file=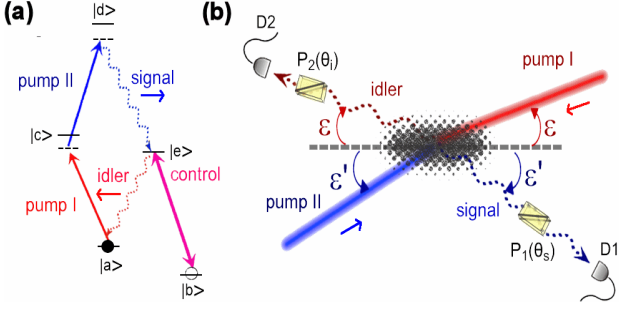,height=1.6in,width=3.3in}
\end{center}
\vspace{-0.7cm} \caption{ (a) The atomic structure for the
  proposed cascade emission scheme involving excitation by
  pumps I and II. Pump II and the signal photons lie in the
  telecommunication wavelength range when a suitable level
  of orbital angular momentum $L=0$ or $L=2$ is used as level $|d\rangle$. For atomic rubidium, the signal wavelength is 1.32
$\mu$m ($6s_{1/2} \rightarrow 5p_{1/2}$ transition),  1.37 $\mu$m ($6s_{1/2} \rightarrow 5p_{3/2}$ transition), 1.48 $\mu$m ($4d_{3/2(5/2)}
\rightarrow 5p_{1/2}$ transition), 1.53 $\mu$m ($4d_{3/2(5/2)} \rightarrow 5p_{3/2}$ transition). For atomic cesium, the signal wavelength is
1.36 $\mu$m ($7s_{1/2} \rightarrow 6p_{1/2}$ transition),  1.47 $\mu$m ($7s_{1/2} \rightarrow 6p_{3/2}$ transition). For Na and K the
corresponding wavelengths are in the 1.1-1.4 $\mu$m range. (b) Schematic of experimental setup based on ultra-cold $^{85}$Rb atomic gas. For
$\lambda_s = 776$ nm, phase matching results in the angles $\varepsilon ^{\prime} \approx \varepsilon \approx 1^{\circ }$, while for $\lambda_s =
1.53$ $\mu$m,  $\varepsilon^{\prime} \approx 2 \varepsilon \approx 2^{\circ }$. $P_1$ and $P_2$ are polarizers; D1 and D2 are detectors.
}\label{TQ}
\end{figure}

The direction of the idler field is determined by the phase matching condition $\vec k_1 +\vec k_2 = \vec k_s + \vec k_i$, where $\vec k_1$ and
$\vec k_2$ are the wavevectors of the laser fields I and II, respectively. Under conditions of phase matching, collective enhancement causes
emission of the the idler photon correlated with a return of the atom into the Zeeman state from which it originated \cite{matsukevich1}. The
fact that the atom begins and ends the absorption-emission cycle in the same state is essential for strong signal-idler polarization
correlations. The reduced density operator for the field, taking into account collective enhancement, was derived in Ref. \cite{jenkins}:
\begin{equation}
  \hat{\rho}(t)  \approx \left(1+ \sqrt{\epsilon}\hat{A}_{2}^{\dag}(t) \right)
  \hat{\rho}_{vac}\left(  1+\sqrt{\epsilon}\hat{A}_{2}(t)  \right),
\end{equation}
where $\hat{\rho}_{vac}$ is the vacuum state of the field, $\hat{A}_{2}^{\dag}(  t)  $ is a time dependent two photon creation operator for the
signal and idler fields, and $\epsilon \ll 1$. For linearly polarized pumps with parallel (vertical) polarizations, we find
\begin{equation}
\hat{A}_{2}^{\dag}(t)  =\cos\chi~\hat{a}_{H}^{\dag}\hat{b}_{H}^{\dag}+\sin\chi\hat{a}_{V}^{\dag}\hat{b}_{V}^{\dag}
\end{equation}
where $\chi$ is determined by Clebsch-Gordan coupling coefficients \cite{jenkins}, $\hat{a}_{H(V)}^{\dag}$ and $\hat{b}_{H(V)}^{\dag}$ are
creation operators for a horizontally (vertically) polarized signal and idler photon, respectively. For the hyperfine level configuration
$F_a=3\rightarrow F_c=4=F_e \rightarrow F_d=5$, and for an unpolarized atomic sample, we find $\sin \chi=2\cos \chi =2/\sqrt{5}$.

 {\it Step (C)} - The photonic qubit is encoded in the idler field polarization. Photonic to atomic qubit conversion was
achieved in Ref.\cite{matsukevich2}. Such conversion can be performed either within the same ensemble or in a suitably prepared adjacent
ensemble/pair of ensembles. In either case, a strong laser control beam is required to couple the other ground hyperfine level $|b \rangle $ to
the intermediate level $|e\rangle $. Collective excitations involving two orthogonal hyperfine coherences serve as the logical states of the
atomic qubit \cite{matsukevich,matsukevich1,matsukevich2}.

{\it Experiment.} We observe phase-matched cascade emission of entangled photon pairs, using samples of cold $^{85}$Rb atoms, for two different
atomic cascades: (a) at $\lambda _s=776$ nm, via the $5s_{1/2} \rightarrow 5d_{5/2}$ two-photon excitation, (b) at $\lambda _s=1.53$ $\mu$m, via
the $5s_{1/2} \rightarrow 4d_{5/2}$ two-photon excitation. The investigations are carried out in two different laboratories using similar setups,
Fig. 1(b). A magneto-optical trap (MOT) of $^{85}$Rb provides an optically thick cold atomic cloud. The atoms are prepared in an incoherent
mixture of the level $|a\rangle$,  which corresponds to the $5s_{1/2},F_{a}=3$ ground level, by means of optical pumping. The intermediate level
$|c\rangle = |e\rangle$ corresponds to the $5p_{3/2},F_c=4$ level of the $D_2$ line at 780 nm, and the excited level $|d\rangle$ represents (a)
the $5d_{5/2}$ level with $\lambda _s = 776$ nm, or (b) the $4d_{5/2}$ level with $\lambda _s = 1.53$ $\mu$m. Atomic level $|b\rangle $
corresponds to $5s_{1/2},F_{b}=2$, and could be used to implement the light-to-matter qubit conversion \cite{matsukevich2}.

The trapping and cooling light as well as the quadrupole magnetic field of the MOT are switched off for the 2 ms duration of the measurement. The
ambient magnetic field is compensated by three pairs of Helmholtz coils. Counterpropagating pumps I (at 780 nm) and II (at 776 nm or 1.53
$\mu$m), tuned to two-photon resonance for the ${|a\rangle \rightarrow |d\rangle }$ transition are focused into the MOT using the off-axis,
counter-propagating geometry of Harris and coworkers \cite{balic}.  This two-photon excitation induces phase-matched signal and idler emission.

With quasi-cw pump fields, we perform photoelectric coincidence detection of the signal and idler fields. The latter are directed onto single
photon detectors D1 and D2. For $\lambda _s=1.53$ $\mu$m, the signal field is coupled into 100 m of single-mode fiber, and detector D1 (cooled
InGaAs photon counting module) is gated using the output pulse of silicon detector D2. The electronic pulses from the detectors are fed into a
time-interval analyzer with 1 ns time resolution.

\begin{figure}[btp]
\begin{center}
\leavevmode  \psfig{file=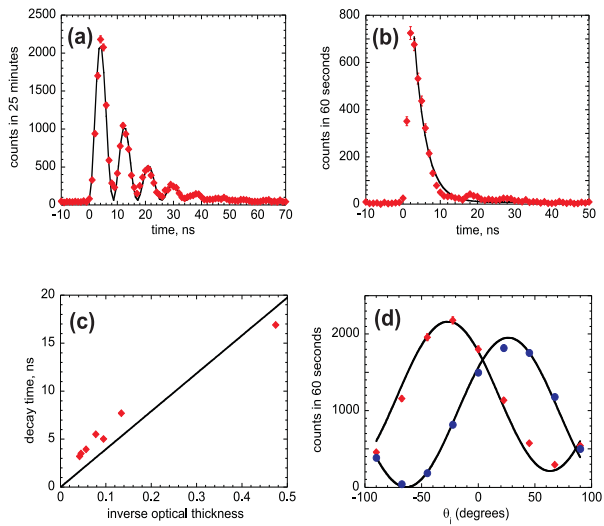,height=3.3in,width=3.3in}
\end{center}
\vspace{-1.6cm} \caption{ (a) Count rate proportional to the signal-idler intensity correlation function $G_{si}$ as a function of signal-idler
delay $\tau$, $|d\rangle = |5d_{5/2}, F=4\rangle$. The quantum beats are associated with 120 MHz hyperfine splitting, $F=3$ and $4$, of the
$5p_{3/2}$ level \cite{vrehen}. The
  solid curve is a fit of the form $\beta+A\exp (-t/\alpha)\sin ^2(\pi \Omega
  t)$, where $\beta=63$,
$A=2972$, $\alpha=11$ ns and $\Omega = 117$ MHz are adjustable
  parameters. (b) Same as (a), but for $|d\rangle = |5d_{5/2}, F=5\rangle$. Since this state can only decay though the $F=4$
component of the $5p_{3/2}$ level, there are no quantum beats. The solid curve is an exponential fit with decay time of $3.2$ ns. (c) The
measured decay time vs the inverse measured optical thickness. (d) Measured coincidence fringes for $\theta _s =45^{\circ }$ (red diamonds) and
$\theta _s =135^{\circ }$ (blue circles). The solid curves are fits based on Eqs.(1,2), with $\cos \chi =1/\sqrt{5}$.} \label{TQ}
\end{figure}
We measure the stationary signal-idler intensity correlation function $ G_{si} (\tau )  = \langle \mathcal{T}:\hat I_s(t) \hat I_i(t+\tau
):\rangle, $ where the notation $\mathcal{T}::$ denotes time and normal ordering of operators, and $\hat{I}_s$ and $\hat{I}_i$ are the signal and
idler intensity operators, respectively \cite{mandel}. Results for (a) $\lambda _s = 776$ nm and (b) $\lambda _s = 1.53$ $\mu$m are presented in
Fig.~2 and Fig.~3, respectively. In particular, the measured correlation functions are shown in Fig.~2(a,b) and Fig.~3(a). The correlation
function shown in Fig.~2(a) exhibits quantum beats due to the two different hyperfine components of the the $5p_{3/2}$ level \cite{vrehen}. The
correlation times are consistent with superradiant scaling $\sim t_s/d_{th}$, Fig.~2(c), where $t_s \approx 27$ ns for the $5p_{3/2}$ level
\cite{eberly}.

In order to investigate polarization correlations of the signal and idler fields, they are passed through polarizers $P_1$ (set at angle $\theta
_s$) and $P_2$ (set at angle $\theta _i$), respectively, as shown in Fig.~1(b). We integrate the time-resolved counting rate over a window
$\Delta T$ centered at the maximum of the signal-idler intensity correlation function $G_{si} (\tau ) $, with (a) $\Delta T=6$ ns for $\lambda
_s=776$ nm, and (b) $\Delta T =1$ ns for $\lambda _s=1.53$ $\mu$m. The resulting signal-idler coincidence rate $C\left({\theta_s},\theta
_i\right)$ exhibits sinusoidal variation as a function of the polarizers' orientations, as shown in Figs.~2(d) and 3(b). In order to verify the
predicted polarization entanglement, we check for violation of Bell's inequality $S\leq 2$ \cite{chsh,mandel,walls}. We first calculate the
correlation function $E\left(\theta _s,\theta _i\right)$, given by
$$
\frac{C\left(\theta _s,\theta _i\right)+C\left(\theta _s^\bot,\theta _i^\bot\right)-C\left(\theta _s^\bot,\theta _i\right)-C\left(\theta
_s,\theta _i^\bot\right)}{C\left(\theta _s,\theta _i\right)+C\left(\theta _s^\bot,\theta _i^\bot\right)+C\left(\theta _s^\bot,\theta
_i\right)+C\left(\theta _s,\theta _i^\bot\right)},
$$
where $\theta ^\bot = \theta + \pi /2$, and $ S=|E\left(\theta _s,\theta _i\right) + E\left({\theta_s}^\prime,\theta _i\right)|+|E\left(\theta
_s,\theta _i^\prime\right)-E\left(\theta _s^\prime,\theta _i^\prime\right)| $.
\begin{figure}[btp]
\begin{center}
\leavevmode  \psfig{file=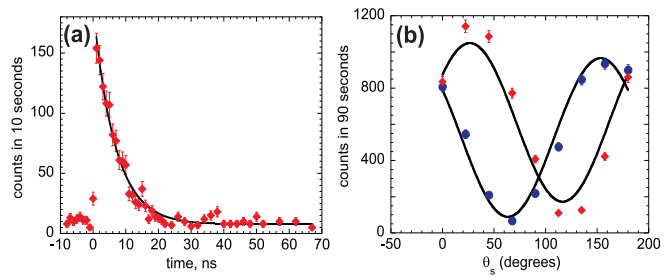,height=1.4in,width=3.4in}
\end{center}
\vspace{-0.7cm} \caption{ (a) Same as Fig.~2(a,b), but for $|d\rangle = |4d_{5/2}, F=5\rangle$. The solid curve is an exponential fit with decay
time of $6.7$ ns. (b) Measured coincidence fringes for $\theta _i =45^{\circ }$ (red diamonds) and $\theta _i =135^{\circ }$ (blue circles). The
solid curves are fits based on Eqs.(1,2), with $\cos \chi =1/\sqrt{5}$.} \label{Fig3}
\end{figure}
\begin{table}
\caption{\label{tab:table1} Measured correlation function $E(\theta _s, \theta _i)$ and $S$  for $\lambda _s =776$ nm and $\lambda _s= 1.53$
$\mu$m.}
\begin{ruledtabular}
\begin{tabular}{ccccc}
$\lambda _s$ &$\theta _s$ & $\theta _i $& $E(\theta_s, \theta _i)$  \\
\hline
&$0^{\circ } $&           $-67.5^{\circ }$     &  $-0.670  \pm 0.011$   \\
&$45^{\circ } $ &      $-22.5^{\circ }$ &  $-0.503  \pm 0.013$   \\
776 nm &$0^{\circ }$  &   $-22.5^{\circ }$     &  $0.577  \pm 0.012$   \\
&  $45^{\circ }$  &    $-67.5^{\circ }$ &  $-0.434 \pm 0.014$   \\
&     &    & $S=2.185 \pm 0.025$           \\
\hline
&  $22.5^{\circ }$  &        $45^{\circ }$ &  $-0.554 \pm 0.027$    \\
&$67.5^{\circ }$  & $0^{\circ }$     &  $-0.682  \pm 0.027$   \\
1.53 $\mu$m &$22.5^{\circ }$  &              $0^{\circ }$     &  $0.473  \pm 0.024$ \\
&$67.5^{\circ }$  &            $45^{\circ }$ &  $-0.423  \pm 0.029$   \\
&     &    & $S=2.132 \pm 0.036$           \\
\end{tabular}
\end{ruledtabular}
\end{table}
\begin{figure}[btp]
\begin{center}
\leavevmode  \psfig{file=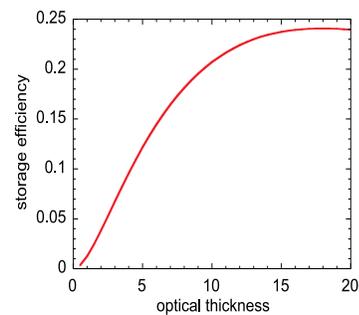,height=1.6in,width=1.8in}
\end{center}
\vspace{-0.7cm} \caption{Efficiency of storage and subsequent retrieval of a coherent idler field with decay time of 6 ns in an auxiliary atomic
ensemble, obtained by numerical integration of the Maxwell-Bloch equations \cite{chaneliere,jenkins,jenkins0}. The atomic coherence time is
assumed to be much longer than the storage time.} \label{Fig4}
\end{figure}

Measured values of $E\left({\theta_s},\theta _i\right)$, using the set of angles $\theta_s,\theta _i$, chosen to maximize the violation of Bell's
inequality, are presented in Table 1. We find (a) $S=2.185 \pm 0.025$ for $\lambda _s=776$ nm, and (b) $S=2.132 \pm 0.036$ for $\lambda _s=1.53$
$\mu$m, consistent with polarization entanglement of signal and idler fields in both cases. The entangled two-photon state of Eqs.(1,2), for
$\sin \chi=2/\sqrt{5}$, has a substantial degree of asymmetry. If oppositely, circularly, polarized pumps I and II were used, the corresponding
two-photon state would be symmetric with $\sin\chi = \cos\chi = 1/\sqrt{2}$ \cite{jenkins}.

The quantum repeater protocol involves sequential entanglement swapping via Hong-Ou-Mandel (HOM) interference followed by coincidence detection
\cite{mandel,duan}. High-visibility HOM interference requires that the signal and idler photon wave-packets have no entanglement in the time or
frequency domains \cite{franson}. This may be achieved with excitation pulses that are far detuned from two-photon resonance and with pulse
lengths much shorter than the superradiant emission time $t_s /d_{th}$ of level $|e \rangle $.

The idler field qubit is naturally suited for conversion into an atomic qubit encoded into the collective hyperfine coherence of levels
$|a\rangle = |5s_{1/2}, F=3\rangle$ and $|b\rangle = |5s_{1/2}, F=2\rangle$. To perform such conversion, either the same or another similar
ensemble/pair of ensembles could be employed \cite{matsukevich2}. A time-dependent control laser field resonant on the $|b\rangle = |5s_{1/2},
F=2\rangle \leftrightarrow |e\rangle = |5p_{3/2}, F=3\rangle$ transition could selectively convert one of the two frequency components of the
idler field, shown in Fig.~2(a), into a collective atomic qubit. Pulsed excitation should be used in order to enable the synchronization of the
idler qubit and the control laser. Numerical simulations show that qubit conversion and subsequent retrieval can be done with good efficiency for
moderate optical thicknesses (Fig.~4), even though the idler field temporal profile is shorter than those employed in Ref. \cite{matsukevich2}
(compare with Fig.~3 in Ref. \cite{chaneliere}).

The basic protocols we have outlined can also be applied to single alkali atom emitters. Similar cascade decays in single atoms were used in
early experiments demonstrating violation of local realism \cite{aspect} and single photon generation \cite{grangier}. For alkali metal atoms, it
is necessary to optically pump the atom into a single Zeeman state, e.g., $m=0$, of level $|a\rangle$. A virtual excitation of a single Zeeman
state of level $|d\rangle$ is created with short laser pulses. Coherent Raman scattering to level $|e\rangle$ results in atom-photon polarization
entanglement. In order to prevent spontaneous decay of the level $|e\rangle$, a control field $\pi$-pulse is applied immediately after the
application of the two-photon excitation, transferring the atomic qubit into the ground state where it could live for a long time. It is
important that the $\pi$-pulse duration is shorter than the spontaneous lifetime of level $|e\rangle$. Two-photon interference and photoelectric
detection of signal photons produced by two remote single atom nodes would result in entanglement of these remote atomic qubits \cite{cabrillo}.
Qubit detection for single atoms can be achieved with nearly unit efficiency and in a time as short as 50 $\mu$s \cite{blinov}. Such high
efficiency and speed lead to the possibility of a loophole-free test of Bell's inequality, for atoms separated by about 30 kilometers. Cascaded
entanglement swapping between successive pairs of remote entangled atomic qubits may be achieved via local coupling of one of the atoms from the
first pair and its neighboring partner from the the following pair \cite{duan1}.

We also point out that the cascade level scheme employed here can be used to convert a telecommunications photon into a near-infrared photon
using four-wave mixing. This could potentially be useful because single-photon detectors for the visible and near-infrared currently have much
higher quantum efficiency, and much lower dark count probability, than practically viable (e.g., InGaAs) detectors used at telecommunication
wavelengths.

In summary, we have proposed a practical telecommunication quantum repeater scheme based on cascade transitions in alkali metal atoms. We have
generated entanglement of a pair of 1.53 $\mu$m and 780 nm photons using an ensemble of ultra-cold rubidium atoms. Combined with our recent
demonstration of light-to-matter qubit conversion \cite{matsukevich2}, the key steps of our proposal have now been taken.

We thank Luis Orozco for fruitful discussions, particularly for pointing out the telecommunications transitions in rubidium \cite{gomez}. We also
thank Prem Kumar for helpful interactions, and Shau-Yu Lan for experimental assistance. We gratefully acknowledge the generous loan of an InGaAs
single photon detector by Henry Yeh and BBN Technologies. This work was supported by the Office of Naval Research, National Science Foundation,
NASA, Alfred P. Sloan Foundation, and Cullen-Peck Chair.

\end{document}